\documentclass[epj,nopacs,final]{svjour}
\usepackage{amsmath,amsxtra,amssymb,latexsym,amscd} 
\usepackage{graphicx}
\usepackage{color}

\begin{document}
\title{Inhomogeneous Josephson junction chains:
a superconducting meta-material for superinductance optimization}

\author{D. V. Nguyen \inst{1,2} \and D. M. Basko \inst{1 \mail{denis.basko@lpmmc.cnrs.fr}}}
\institute{ 
Laboratoire de Physique et Mod\'elisation des Milieux Condens\'es,
Universit\'e de Grenoble-Alpes and CNRS,
25 Avenue des Martyrs, 38042 Grenoble, France 
\and  
Institut N\'eel, Universit\'e de Grenoble-Alpes and CNRS,
25 Avenue des Martyrs, 38042 Grenoble, France
}

\authorrunning{D. V. Nguyen and D. M. Basko}
\titlerunning{Inhomogeneous Josephson junction chains for superinductance optimization}


\abstract{
We report a theoretical study of the low-frequency impedance of
a Josephson junction chain whose parameters vary in space.
Our goal is to find the optimal spatial profile which
maximizes the total inductance of the chain without shrinking
the low-frequency window where the chain behaves as an inductor.
If the spatial modulation is introduced by varying the junction
areas, we find that the best result is obtained for
a spatially homogeneous chain, reported earlier in the literature.
An improvement over the homogeneous result can be obtained by
representing the junctions by SQUIDs with different loop areas,
so the inductances can be varied by applying a magnetic field.
Still, we find that this improvement becomes less important for
longer chains.
}

\maketitle

\section{Introduction}

Quantum engineering in superconducting nanocircuits is a
rapidly developing field, due to progress in sample fabrication
techniques which has been occurring in the past decade~\cite{Jung2014}
Complex circuits with many elements can be routinely
fabricated on a chip nowadays. Due to
superconductivity, electromagnetic signals propagate in such
circuits with extremely low losses, and the circuit properties can
be tuned by applying an external magnetic field.
Highly inductive elements are often needed in such nanocircuits,
to realize a large non-dissipative impedance.
Applications of large inductances include protection of fluxonium
qubits from the charge noise~\cite{Manucharyan2009},
tunable microwave impedance matching~\cite{Altimiras2013}, or
a potential implementation of the electrical current standard
in quantum metrology based on Bloch oscillations 
\cite{Likharev1985,Mooij2006,Guichard2010}.

Because any geometrical inductor (a coil being the standard textbook
example) also necessarily possesses a parasitic self-capacitance
which starts to dominate at high frequencies, its non-dissipative
impedance is limited by the vacuum impedance,
$\sim\sqrt{\mu_0/\epsilon_0}=4\alpha R_Q$, where
$\alpha\approx{1}/137$ is the fine structure constant, and
$R_Q\approx 13\:\mathrm{k}\Omega$ is the resistance
quantum~\cite{Feynman}.
Indeed, the inductance of a geometrical inductor is due to the
magnetic field produced by the current, which acts on the current
itself. The relativistic nature of this effect is the intrinsic
reason for its weakness.
This limitation can be overcome by using superconducting
materials whose inductance is due to the kinetic energy of the Cooper
pair condensate~\cite{Tinkham}, and thus is of non-relativistic
origin. The term ``superinductance'' is often used to denote such
superconductivity-based inductance.


Several structures, based on Josephson junctions (JJs), have
been reported to work as superinductors~\cite{Masluk2012,Bell2012}.
In the first one, a large inductance was obtained by putting
$N$~Josephson junctions in series, which gave the total
inductance~$NL$ ($L$~is the inductance of a single junction).
In Ref.~\cite{Bell2012}, magnetic-field-induced frustration
was used to increase the inductance, which then exhibited a
strong nonlinearity. Here, we focus on the linear case, and
analyze structures analogous to that of Ref.~\cite{Masluk2012}.

A simple strategy to increase the total inductance of a JJ~chain
would then be to make $L$ and/or $N$ as large as possible. However,
in either case one faces some limitations. In the first case,
the JJ inductance~$L$ is inversely proportional to the Josephson
energy of the junction, $E_J=(\hbar/2e)^2(1/L)$. To work as an
inductor, the junction must be in the superconducting regime,
$E_J\gg{E}_C$, where the charging energy $E_C=(2e)^2/(2C)$ is
determined by the junction capacitance. This condition sets a
lower limit on~$E_J$, or, equivalently, an upper limit
$L<L_\mathrm{max}$, or a lower limit on the junction
area~$\mathcal{A}$, as both $E_J,C\propto\mathcal{A}$.

\begin{figure}
\centering
\includegraphics[width=7cm]{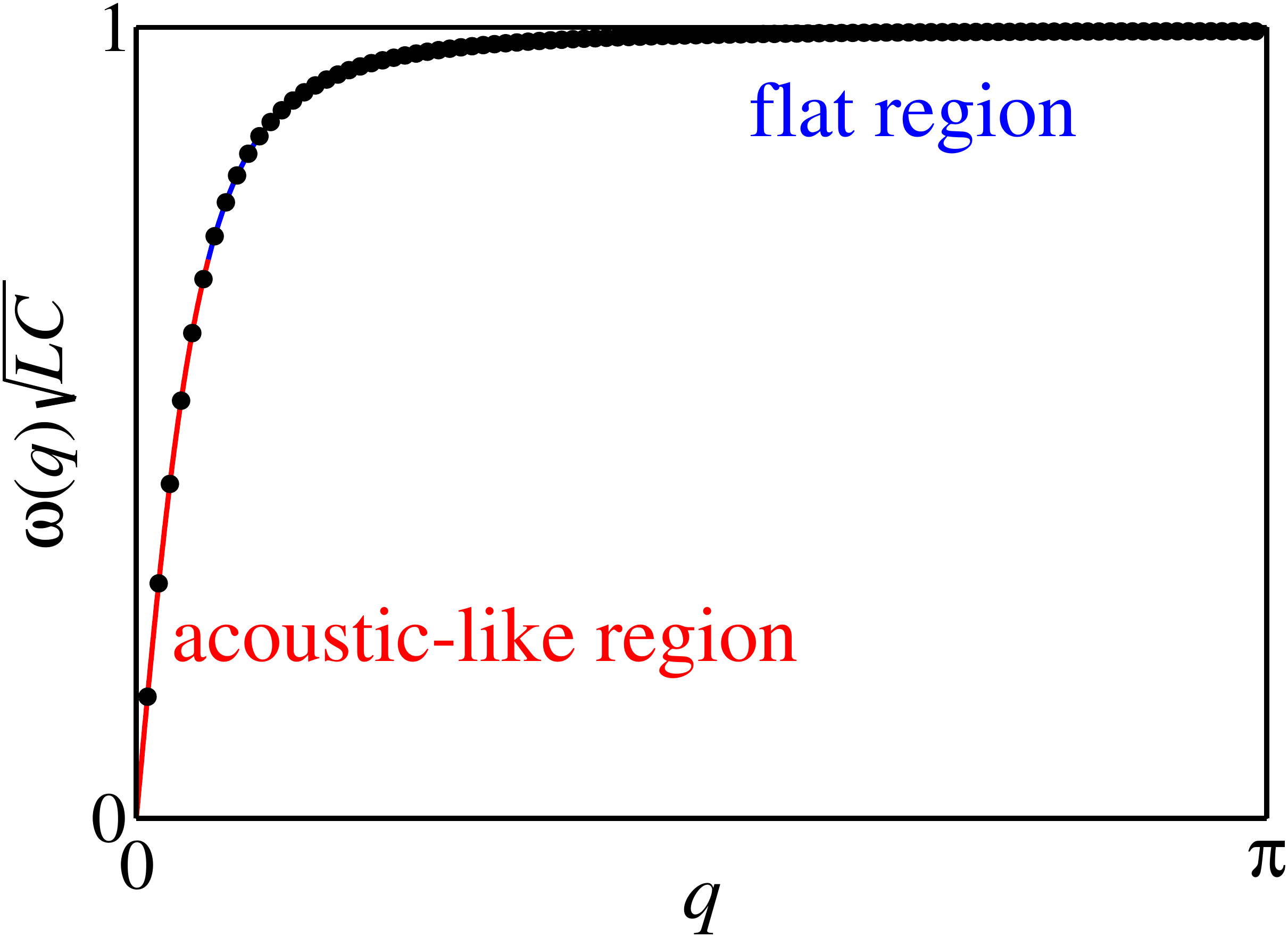}
\caption{\label{fig:dispersion}
Dispersion curve of a JJ chain with $C/C^\mathrm{g}=25$
(solid curve) and modes of a chain with $N=100$ junctions
(filled circles).
The mode frequency $\omega$ is measured in the units of the
junction plasma frequency $1/\sqrt{LC}$.
}
\end{figure}

Limitations on the junction number~$N$ arise from the
dependence of the chain response on the frequency~$\omega$.
The phase slip rate, although exponentially suppressed for
$E_J\gg{E}_C$, grows with~$N$, giving rise to a finite dc
resistance, which spoils the purely inductive response of the
chain at low frequencies. From the high-frequency side, the
effective bandwidth of the inductive response is restricted
by electromagnetic modes supported by the chain,
$\omega\ll\omega_1$ (the lowest mode frequency).
Crucially, besides the capacitance~$C$ of the junction between
neighboring superconducting islands, each island has a small
capacitance~$C^\mathrm{g}$ to the ground. This capacitance
gives rise to screening of the Coulomb interaction between the islands
on a length scale $\lambda=\sqrt{C/C^\mathrm{g}}$ and produces an
acoustic-like region of the mode dispersion $\omega(q)=(LC)^{-1/2}\sqrt{\epsilon^2(q)/[\epsilon^2(q)+\lambda^{-2}]}$ of 
spatially homogeneous chains, where $\epsilon(q)=2\sin(q/2)$, and $q$~is the wavenumber,
$0\leqslant{q}\leqslant\pi$ (Fig.~\ref{fig:dispersion}).
The first mode corresponds to $q=\pi/(N+1)$, so for large
$N\gg\pi\lambda$, the
frequency of the lowest mode $\omega_1\propto{1}/N$, and the
inductive response bandwidth shrinks with increasing~$N$. This
was the main limitation for the device studied in
Ref.~\cite{Masluk2012}, where a special effort was made
to decrease the parasitic ground capacitance~$C^\mathrm{g}$.

The above argumentation works for spatially homogeneous
chains, whose total inductance is determined by just two
parameters, the single-junction inductance~$L$ and their
number~$N$, if $L$~is assumed to be the same for all
junctions. This, however, need not be the case, since an
arbitrary spatial profile of junction sizes along the
chain can be produced during the sample fabrication.
A spatial modulation of junction parameters modifies the
normal modes of the chain, and can manifest itself in
various situations.
For example, Josephson energy renormalization by coupling
to the normal modes was shown to be affected by a modulation
of the chain parameters~\cite{Taguchi2015}.
Effect of the normal mode structure on dephasing of the
fluxonium qubit was discussed in Ref.~\cite{Viola2015}.
For the present problem, one can try to optimize
the total inductance and the operation bandwidth of the chain
using many more degrees of freedom than just $L$~and~$N$,
because the parameters of each of the $N$~junctions
can be treated as optimization variables. To study,
whether one can take advantage of this large number of
variables and improve the homogeneous chain
result of Ref.~\cite{Masluk2012} by carefully choosing
the spatial profile of the junction parameters, is the
purpose of the present work.

In this paper, we consider two ways to introduce a spatial
inhomogeneity into the structure.
One is to vary the area $\mathcal{A}_n$ of each junction~$n$
(assuming the island area to be already optimized to minimize
the ground capacitance as was done in
{Refs.~\cite{Manucharyan2009,Masluk2012,Manucharyan2012}}).
This leads to a simultaneous variation of the junction
inductances $L_n$ and capacitances~$C_n$, such that their
product $L_nC_n=\mathrm{const}$. Optimizing over all areas
$\{\mathcal{A}_n\}$, we find that the best result is still
achieved for a homogeneous configuration.

\begin{figure}
\centering
\includegraphics[width=7cm]{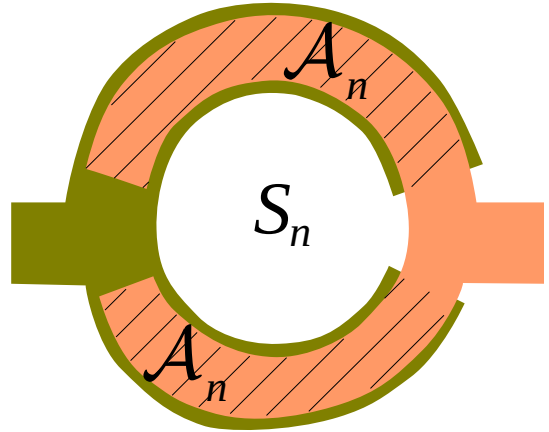}
\caption{\label{fig:SQUID}
A schematic representation of a SQUID (top view):
two superconducting islands, top and bottom, are connected
by two junctions forming a loop. At zero magnetic field, the
SQUID inductance $L_n(0)$ is determined by the total area of
the junctions~$\mathcal{A}_n$, shown by hatching. When a magnetic
field~$B$ is applied, the inductance
$L_n(B)=L_n(0)/|\cos(\pi{B}S_n/\Phi_0)|$ is determined by
the magnetic flux $BS_n$ through the SQUID loop area~$S_n$,
represented by the white circular region in the center.
}
\end{figure}

The second way to introduce a spatial variation of the junction
parameters is to represent each junction by a SQUID
(superconducting quantum interference device). When subject
to a magnetic field~$B$, a SQUID behaves like an effective
Josephson junction with a field-dependent Josephson energy
$E_J(B)=E_J(0)|\cos(\pi{B}S/\Phi_0)|$, where
$\Phi_0=2\pi\hbar/(2e)$ is the superconducting flux quantum,
and $S$~is the SQUID loop area which determines the magnetic
flux $BS$ through the SQUID (Fig.~\ref{fig:SQUID}).
Then, if all SQUIDs have different
areas $S_n$, the inductance of each junction of the chain,
$L_n(B)=L_n(0)/|\cos(\pi{B}S_n/\Phi_0)|$, varies in space,
and this variation is independent of the variation of the
capacitance~$C_n$ (the latter is controlled by the junction
area~$\mathcal{A}_n$, independent of the loop area~$S_n$).
In this case, we show that one can indeed improve over the
homogeneous result, by placing SQUIDs with larger loop area
(higher inductance) near the ends of the chain. Still, the
obtained improvement over the homogeneous result turns out
to decrease with the increasing chain length.

The paper is organized as follows.
In the next section we specify the model and formally pose
the optimization problem.
In Sec.~\ref{sec:junctionarea} we analyze the case when only
the junction areas~$\mathcal{A}_n$ vary in space.
In Sec.~\ref{sec:looparea} we study variation of the SQUID
loop areas~$S_n$.
In Sec.~\ref{sec:conclusions} we give our conclusions.

\section{Formal setting of the optimization problem}
\label{sec:Model}

\begin{figure}
\includegraphics[width=8cm]{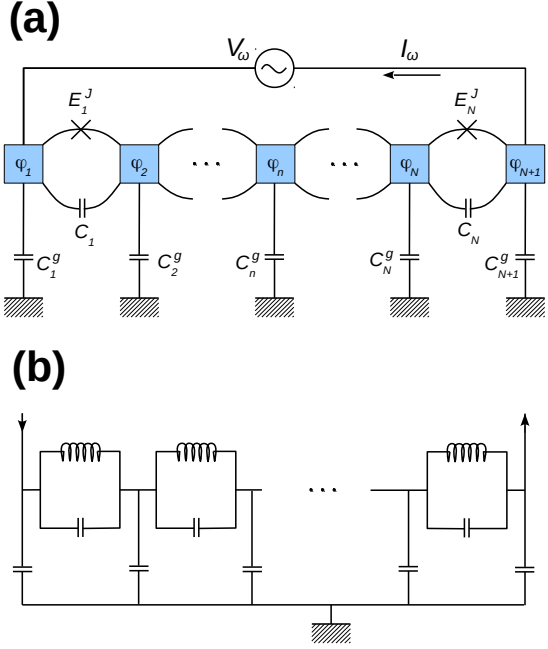}
\caption{\label{fig:JJchain}
(a) A schematic view of the Josephson junction chain and its
impedance definition.
(b)~Linear circuit, equivalent to the chain shown in~(a),
described by Eqs.~(\ref{Kirchhoff=}).}
\end{figure}

We consider a chain of $N+1$~superconducting islands. Each
island is connected to its nearest neighbors by Josephson
junctions, so the chain has $N$~junctions
(Fig.~\ref{fig:JJchain}). We assume $N\gg{1}$.
When the junctions are in the
superconducting regime, $E_J\gg{E}_C$, the oscillations of the
superconducting phase $\varphi_n$ on each island are small.
Then, the Josephson current through the $n$th junction from
island~$n$ to~$(n+1)$ can be written as
$I_{n}=I_{n}^c\sin(\varphi_{n+1}-\varphi_n) \approx
I_{n}^c(\varphi_{n+1}-\varphi_n)$~\cite{Tinkham}.
Here, $I_{n}^c=\hbar/(2eL_n) = (2e/\hbar)E_{J,n}$ is the
junction critical current. Thus, the voltage drop across the
junction can be determined by using the Josephson relation:
\begin{equation}
V_{n+1}-V_n = \frac\hbar{2e}
\left(\frac{d\varphi_{n+1}}{dt} - \frac{d\varphi_n}{dt}\right)
= \frac\hbar{2eI_{n}^c}\frac{dI_n}{dt}. 
\end{equation}
This expression shows that
the junction behaves as a linear inductor, and the Josephson kinetic
inductance is given by $L_{n}=\hbar /(2eI_{n}^c)$. In addition,
we assume that the dissipation is very small and can be
neglected. Then, an isolated chain is equivalent to the electric
circuit shown in Fig.~\ref{fig:JJchain}(b), where $C_{n}$ is
the capacitance formed by the neighboring superconducting
islands, and $C^\mathrm{g}_{n}$ is the capacitance of each
island to ground.
We define the complex impedance $Z(\omega)$ of the chain at
frequency~$\omega$ as the ratio of the voltage
$V_\omega{e}^{-i\omega{t}}$ on an external ac voltage source,
connected to the islands $n=1$ and $n=N+1$,
to the current $I_\omega{e}^{-i\omega{t}}$ through this source
(Fig.~\ref{fig:JJchain}).

To determine the normal mode frequencies of this circuit, one can apply
the Kirchhoff's law at each of the $N+1$ nodes of this circuit.
This gives the following system of linear equations for the voltages~$V_n$:
\begin{subequations}\label{Kirchhoff=}\begin{eqnarray}
\label{eq:Kirchhoff1}
&& Y_{1}({V_1} - {V_2})-i\omega{C}^\mathrm{g}_1V_1 = I_\omega,   \\
\label{eq:Kirchhoff2}
&& Y_n(V_n-V_{n+1})+Y_{n-1}(V_n - V_{n - 1})-i\omega{C}^\mathrm{g}_nV_n = 0
\nonumber\\ &&\hspace*{5cm} (n=2,\dots,N),\\
\label{eq:KirchhoffN}	
&& Y_N(V_{N+1} - V_N) -i\omega{C}^\mathrm{g}_{N+1}V_{N+1} = -I_\omega,
\end{eqnarray}\end{subequations}
where the junction admittance is defined as
\begin{equation}
Y_n(\omega) = - i\omega {C_{n}} - \frac{1}{i\omega{L_n}}.
\end{equation}
System~(\ref{Kirchhoff=}) can be written in the matrix form,
$\hat\Upsilon(\omega)\mathbf{V}=\mathbf{I}$, in terms of the column
vectors $\mathbf{V}^T=(V_1,\ldots,V_{N+1})$ and
$\mathbf{I}^T=(I_\omega,0,\ldots,0,-I_\omega)$, as well as
the corresponding matrix $\hat\Upsilon(\omega)$.
Then the chain impedance $Z(\omega)$ can be expressed in terms of
the matrix elements of the inverse $\hat\Upsilon^{-1}(\omega)$ as
$Z=(\hat\Upsilon^{-1})_{11}-(\hat\Upsilon^{-1})_{1,N+1}
-(\hat\Upsilon^{-1})_{N+1,1}+(\hat\Upsilon^{-1})_{N+1,N+1}$.
At low frequencies, the admittances are dominated by the inductive
part, so the impedance is given by
$Z(\omega\to{0})=-i\omega{L}_\mathrm{tot}$, where $L_\mathrm{tot}$ is the
total inductance of the chain,
\begin{equation}\label{Ltot=}
L_\mathrm{tot}=\sum_{n=1}^NL_n.
\end{equation}
The approximation $Z(\omega)\approx-i\omega{L}_\mathrm{tot}$ is valid as
long as $\omega\ll\omega_1$, where $\omega_1$ is the lowest normal
mode frequency, for which $\det\hat\Upsilon(\omega)=0$.

As discussed in the introduction, ideally one would like to increase
both $L_\mathrm{tot}$ and $\omega_1$, but these two requirements are in
conflict. Thus, one can try to maximize $L_\mathrm{tot}$ at
fixed~$\omega_1$, or maximize $\omega_1$ while keeping $L_\mathrm{tot}$
fixed. We prefer the second option, as the constraint expressed by
Eq.~(\ref{Ltot=}) is much easier to resolve than the constraint
$\omega_1=\mathrm{const}$. Thus, our optimization problem is formulated
as follows: find the spatial profile of $L_n,C_n,C_n^\mathrm{g}$ which
maximizes $\omega_1$ while keeping $L_\mathrm{tot}$ fixed.
To complete the formulation of the problem, we have to specify the
independent variables over which the optimization is performed.

The shape and size of the superconducting islands and of the
junctions between them
can be well controlled in the fabrication process. It is easy to
notice that while the parameters $L_n,C_n$ are mostly determined
by the junction areas, the parasitic ground capacitances
$C_n^\mathrm{g}$ are mostly determined by the island sizes. Thus,
the first obvious step is to minimize the island sizes as much as
possible while keeping constant the junction areas, as any part of
the island area which does not participate in the junctions, does
not contribute to the inductance, but decreases $\omega_1$. This
optimization was performed in
Refs.~\cite{Manucharyan2009,Masluk2012,Manucharyan2012,ManucharyanThesis}.
Then, the ground capacitance of the $n$th island becomes a
function of the areas of the junctions in which it particpates,
$n-1$ and $n$. This function was calculated numerically in
Ref.~\cite{Masluk2012}, and the resulting dependence resembles
a weak power law or a logarithm. We will assume that this first
optimization step has been performed.

Then, our first setting corresponds to independent variation of
all junction areas, which are allowed to vary in a certain range.
In the fabrication process, quite a wide range of sizes can be
achieved, and the restriction on the areas rather comes from
physical considerations. 
One restriction is that for too small areas,  the condition
$E_J\gg{E}_C$ is violated, and then the classical description of
small phase oscillations is no longer valid. Indeed, the amplitude
of a quantum phase slip,
$\propto{e}^{-(2/\pi)R_Q/Z_n}$~\cite{Matveev2002} is exponentially
suppressed only for small junction impedances,
$Z_n\equiv \sqrt{L_n/C_n}\ll{R}_Q$
(we remind that $R_Q\approx 13\:\mathrm{k}\Omega$ denotes the
resistance quantum), so too large impedances are not allowed.
The junction impedance is inversely proportional to its area,
so the area cannot be made too small. On the other hand,
if the junction area is too large, the junction can no longer be
treated as a zero-dimensional object, because the frequency of
its own electromagnetic modes becomes too low. 

Let us choose the smallest allowed junction area as the unit
of area. Then the largest allowed junction area
$\mathcal{A}_\mathrm{max}\gg{1}$ is an independent dimensionless
parameter of the problem. The junction inductance and capacitance
at the smallest area, $L_\mathrm{max}$ and $C_\mathrm{min}$,
can be chosen as the units of inductance and capacitance,
respectively.
Thus, we have
$N$ dimensionless variables $\mathcal{A}_n$, allowed to vary in
the range
\begin{subequations}
\begin{equation}\label{Ahypercube=}
1\leqslant\mathcal{A}_n\leqslant\mathcal{A}_\mathrm{max}.
\end{equation}
They determine the inductance and the capacitance of each junction as
\begin{equation}\label{LnCn=An}
L_n=\frac{L_\mathrm{max}}{\mathcal{A}_n},\quad
C_n=C_\mathrm{min}\mathcal{A}_n,
\end{equation}
and Eq.~(\ref{Ltot=}) thus imposes a constraint on the set $\{\mathcal{A}_n\}$. 
{In this case, the plasma frequency of each junction is unchanged, $1/\sqrt {{L_n}{C_n}}  = 1/\sqrt {{L_{\max }}{C_{\min }}}  \equiv {\omega _p}$}.
Finally, for the ground capacitances we use a simple form
\begin{equation}\label{Cgn=An}
C^\mathrm{g}_n=C^\mathrm{g}_\mathrm{min}\,g(\mathcal{A}_{n-1}/2+\mathcal{A}_n/2),
\end{equation}
\end{subequations}
where $g(x)$ is some function, growing sublinearly with~$x$
(a power law or a logarithm). All qualitative arguments given
below are not sensitive to the specific dependence~$g(x)$;
in the numerical calculations, we set $g(x)=\sqrt{x}$,
{as mentioned in Ref.~\cite{ManucharyanThesis}.}
To define Eq.~(\ref{Cgn=An}) at the ends, we set
$\mathcal{A}_0\equiv\mathcal{A}_1$,
$\mathcal{A}_{N+1}\equiv\mathcal{A}_N$.
Thus, the first optimization problem is fully defined as
maximization of $\omega_1$ determined from Eqs.~(\ref{Kirchhoff=}),
whose coefficients are expressed by Eqs.~(\ref{LnCn=An})
and~(\ref{Cgn=An}) in terms of the dimensionless areas~$\mathcal{A}_n$.
The optmization variables are the areas $\mathcal{A}_n$ in the allowed
range~(\ref{Ahypercube=}) and subject to constraint~(\ref{Ltot=}),
\emph{as well as} the number of the junctions~$N$ itself.
Note that constraint~(\ref{Ltot=}) and inequalities~(\ref{Ahypercube=})
restrict the number of junctions~$N$ to the interval
\begin{equation}
N_0\equiv\frac{L_\mathrm{tot}}{L_\mathrm{max}}\leqslant{N}
\leqslant{N}_0\mathcal{A}_\mathrm{max}.
\end{equation}

The second way of producing a spatial variation of the JJ chain
parameters is to replace each junction by a SQUID. Each SQUID is
characterized by its loop area~$S_n$, independent of the junction
area~$\mathcal{A}_n$ (Fig.~\ref{fig:SQUID}). By applying a magnetic field~$B$,
one can change the SQUID inductance as
\begin{equation}
L_n(B)=\frac{L_n(0)}{|\cos(\pi{B}S_n/\Phi_0)|}
\end{equation}
where the zero-field inductance $L_n(0)$ is determined by the
junction area~$\mathcal{A}_n$. 
This way
of tuning the properties of the JJ by magnetic field is routinely
used in experiments (see, e.~g., Ref.~\cite{Weissl2015}).
Here, it is crucial for us
that the spatial variation of inductance is independent of that
of capacitance, which was not the case in the previous model,
since in Eq.~(\ref{LnCn=An}) the product $L_nC_n$ remained fixed.
Thus, instead of the optimization problem defined by
Eqs.~(\ref{Ahypercube=})--(\ref{Cgn=An}) via variables
$\mathcal{A}_1,\ldots,\mathcal{A}_N$,
we consider another problem defined via variables
$\mathcal{F}_1,\ldots,\mathcal{F}_N$:
\begin{subequations}\begin{align}\label{Fhypercube=}
&1\leqslant\mathcal{F}_n\leqslant\mathcal{F}_\mathrm{max},\\
&L_n=\frac{L_\mathrm{max}}{\mathcal{F}_n},\quad C_n=C_\mathrm{min},\quad
C_n^\mathrm{g}=C^\mathrm{g}_\mathrm{min},\label{Ln=Fn=}
\end{align}\end{subequations}
All junction areas are assumed to be the same, $\mathcal{A}_n=1$, 
{so the plasma frequency of each SQUID is modulated as
$1/\sqrt{L_nC_n}  = {\omega_p}\sqrt{\mathcal{F}_n}$},
and each variable $\mathcal{F}_n$ represents the ratio
\begin{equation}
\mathcal{F}_n=\frac{|\cos(\pi{B}S_n/\Phi_0)|}%
{\cos(\pi\Phi_\mathrm{max}/\Phi_0)},\quad
\frac{1}{\mathcal{F}_\mathrm{max}}\equiv
\cos\frac{\pi\Phi_\mathrm{max}}{\Phi_0},
\end{equation}
where $\Phi_{\mathrm{max}}$ is some maximal magnetic flux allowed to pierce the SQUID loops in order for the device to remain in the superconducting regime $E_J\gg{E}_C$.
Clearly, $\{\mathcal{F}_n\}$ are independent variables,
because $\{S_n\}$ are independent, and additional freedom
is introduced by the magnetic field. Just like before, the
only constraint on $\mathcal{F}_n$ is Eq.~(\ref{Ltot=}),
and it restricts the chain length~$N$ to the interval
\begin{equation}\label{Ninterval=}
N_0\equiv\frac{L_\mathrm{tot}}{L_\mathrm{max}}\leqslant{N}
\leqslant\mathcal{F}_\mathrm{max}N_0.
\end{equation}

The two optimization problems, defined by
Eqs. (\ref{Ahypercube=})--(\ref{Cgn=An}) and by
Eqs. (\ref{Fhypercube=})--(\ref{Ln=Fn=}),
will be studied in the next two sections, respectively.

\section{Junction area modulations}
\label{sec:junctionarea}
Before we proceed with optimization for inhomogeneous JJ chains,
it is useful to see what can be achieved in the homogeneous case,
for future reference. For the problem
(\ref{Ahypercube=})--(\ref{Cgn=An}), with all
$\mathcal{A}_n=\mathcal{A}$, we have only two variables,
$\mathcal{A}$ and $N$. Constraint~(\ref{Ltot=})
fixes $\mathcal{A}=N/N_0$, $L_n=L_\mathrm{max}N_0/N$,
$C_n=C_\mathrm{min}N/N_0$,
$C^\mathrm{g}_n=C^\mathrm{g}_\mathrm{min}\,g(N/N_0)$.
It is convenient to denote the first mode frequency for
this homogeneous chain by~$\Omega_N$. It is given by~\cite{Masluk2012}
\begin{align}\label{OmegaNA=}
\Omega_N^2={}&{}\frac{1}{LC}\,
\frac{1-\cos[\pi/(N+1)]}{1-\cos[\pi/(N+1)]+C^\mathrm{g}/(2C)}
\approx\nonumber\\
\approx{}&{}\frac{(L_\mathrm{max}C_\mathrm{min})^{-1}}%
{1+(C^\mathrm{g}_\mathrm{min}/C_\mathrm{min})(N_0/\pi)^2[x/g(x)]^2}
\end{align}
for $N\gg{1}$.
This is a decreasing function of $x\equiv{N}/N_0$ for any
$g(x)$ growing slower than linearly with~$x$. Thus, $\omega_1$ is
maximized by taking $N=N_0$, all $L_n=L_\mathrm{max}$. We denote
the corresponding value of $\omega_1$ by $\Omega_{N_0}$.

To improve this result using an inhomogeneous chain, one should
take some~$N>N_0$ [a smaller one would be incompatible with
the constraint~(\ref{Ltot=})], and hope that the gain in~$\omega_1$
from the inhomogeneiety would overcome the loss due to the length
increase.
A qualitative idea of the best spatial profile~$\mathcal{A}_n$ can
be obtained from the perturbation theory for system~(\ref{Kirchhoff=}),
developed in Ref.~\cite{Basko2013}.
Let us use the homogeneous chain of length $N$ with all
$\mathcal{A}_n=\mathcal{A}=N/N_0$ and the first mode
frequency $\Omega_N$ as the zero approximation.
If we now modify each junction area by a small amount
$\Delta\mathcal{A}_n$, the first-order frequency shift
is given by~\cite{Basko2013}
\begin{subequations}\begin{align}\label{perturbative=}
&\frac{\Delta\omega_1}{\Omega_N}=
\frac{1}{N+1}\left(1-\Omega_N^2L_\mathrm{max}C_\mathrm{min}\right)
\sum\limits_{n=1}^N\alpha_n\,\frac{\Delta\mathcal{A}_n}{\mathcal{A}},\\
&\alpha_n=\sin^2\frac{\pi{n}}{N+1}+
\frac{2{\mathcal{A}}g'(\mathcal{A})}{g(\mathcal{A})}\times{}\nonumber\\
&\qquad{}\times\left(\cos\frac\pi{N+1}\sin^2\frac{\pi{n}}{N+1}
-\cos^2\frac{\pi/2}{N+1}\right).
\end{align}\end{subequations}
The dependence of $\alpha_n$ on $n$ is quite simple
($\sin^2+\:\mathrm{const}$), and
$\alpha_n$ is the largest for $n=(N+1)/2$, in the middle of the
chain. The value at the maximum $\alpha_{(N+1)/2}>0$ as long as
$[2\mathcal{A}g'(\mathcal{A})/g(\mathcal{A})]\sin^2[\pi/(N+1)]<1$,
which is the case for any sublinear $g(x)$ and $N>4$.
Thus, the center of the chain
contributes the most to the increase of $\omega_1$.

Let us take $N=N_0+1$. Then, the largest increase of the areas
near the center, allowed by constraint~(\ref{Ltot=}), is obtained
by keeping $N_0-1$ junctions with $\mathcal{A}_n=1$, and two more junctions
with $\mathcal{A}_n=2$, to be put in the center. (Note that it is impossible
to keep $N_0$~junctions with $\mathcal{A}_n=1$, as the constraint would
require the remaining one to have $\mathcal{A}_n=\infty$). As the area
change for the central junctions is not small, the perturbative
Eq.~(\ref{perturbative=}) is not sufficient to describe this
situation. Still, $\omega_1$ for this structure can be found
analytically. The result of this straightforward but bulky
calculation, given in \ref{app:2central}, is that the
resulting frequency \emph{is always smaller} than~$\Omega_{N_0}$.

\begin{figure}
\includegraphics[width=8cm]{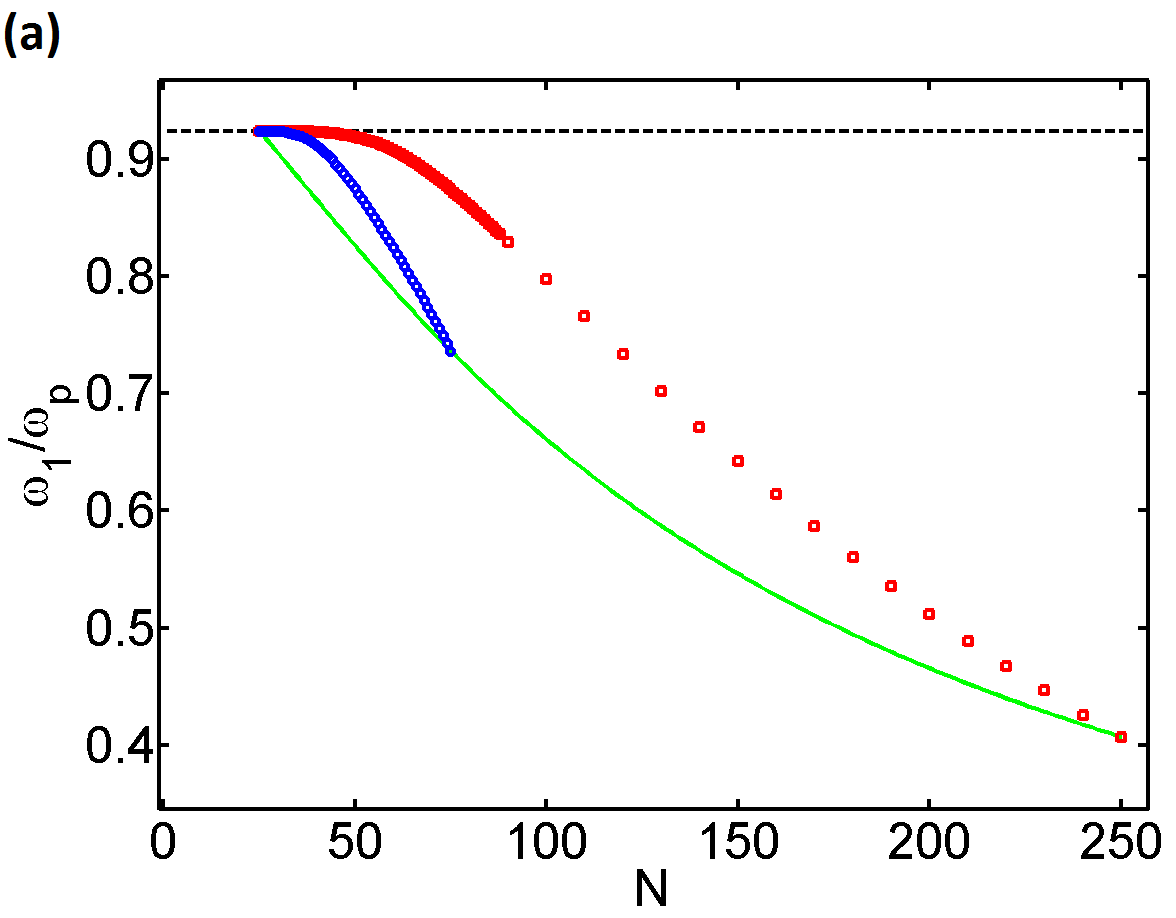}
\includegraphics[width=8cm]{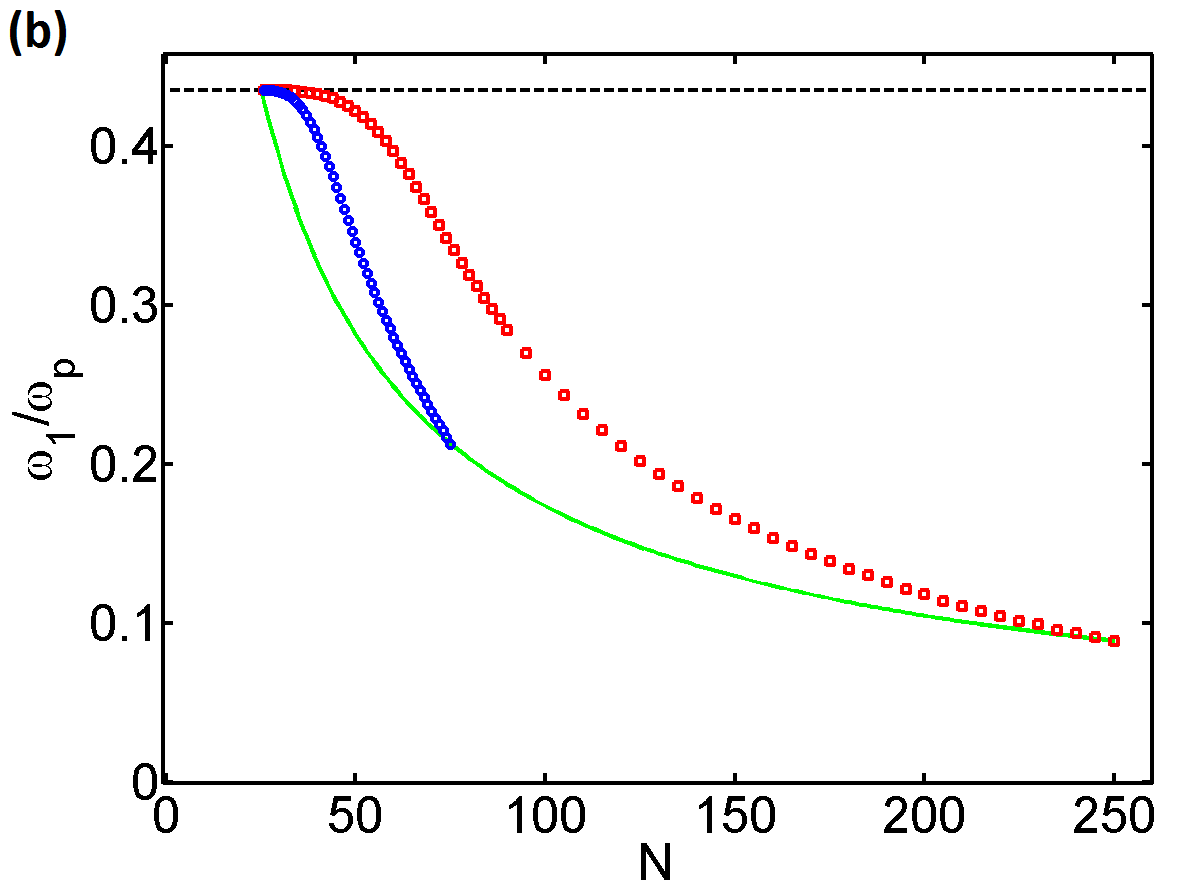}
\caption{\label{fig:areaomega1}
The first mode frequency~$\omega_1$ (in units of the plasma
frequency
$\omega_\mathrm{p}\equiv{1}/\sqrt{L_\mathrm{max}C_\mathrm{min}}$)
obtained by full numerical optimization of all junction areas
$\{\mathcal{A}_n\}$, subject to constraint~(\ref{Ltot=}).
We take $N_0=25$ for all curves, while
$\lambda^2\equiv
C_\mathrm{min}/C^\mathrm{g}_\mathrm{min}=400$ and~16 for panels
(a) and (b), respectively.
Two values of $\mathcal{A}_\mathrm{max}=3$ and 10 were chosen,
shown by the blue and red symbols (lower and upper curves),
respectively, on each panel.
The solid curve shows $\Omega_N$, the first mode frequency for
the homogeneous chain with $L_n=L_\mathrm{max}N_0/N$,
$C_n=C_\mathrm{min}N/N_0$,
$C^\mathrm{g}_n= C^\mathrm{g}_\mathrm{min}\sqrt{N/N_0}$,
and the dashed horizontal line shows the best homogeneous
result~$\Omega_{N_0}$.
}
\end{figure}

\begin{figure}
\centering
\includegraphics[width=8cm]{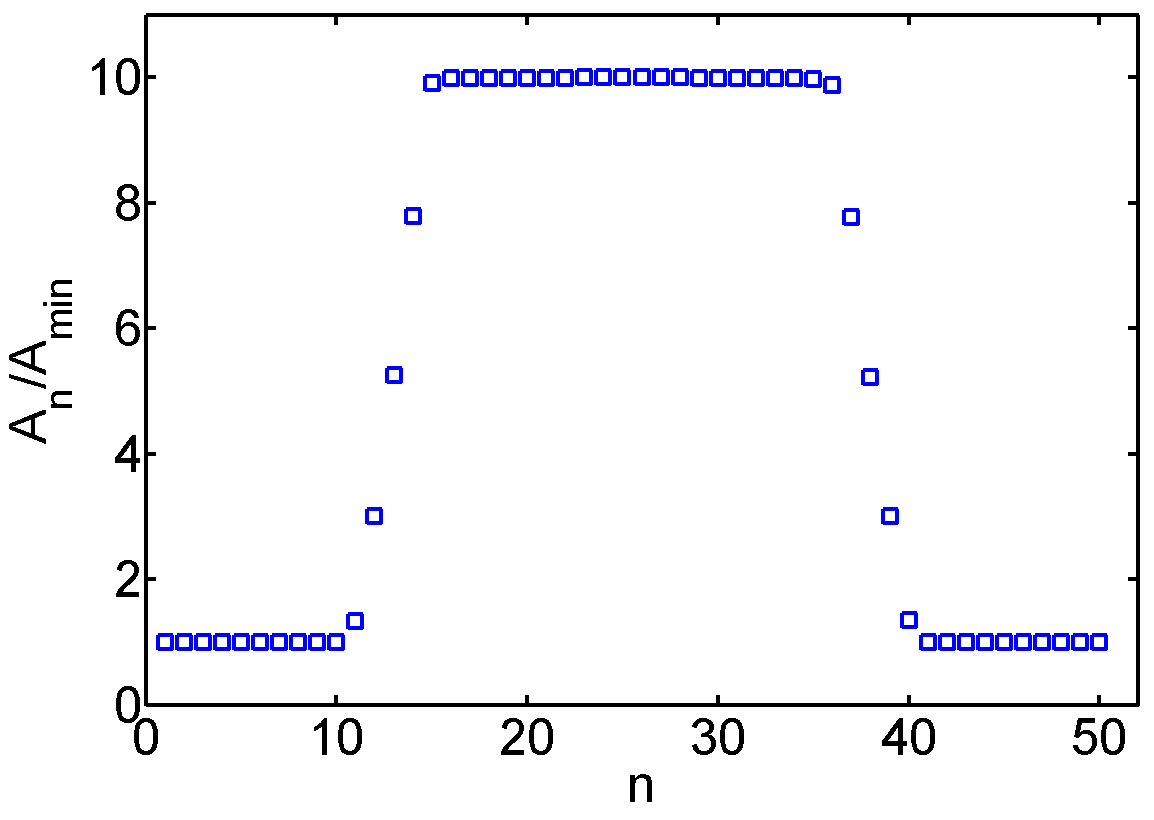}
\caption{\label{fig:areaprofile}
The optimal spatial profile $\{\mathcal{A}_n\}$, giving the largest $\omega_1$
for $N_0=25$, $N=50$, $\mathcal{A}_\mathrm{max}=10$,
$C_\mathrm{min}/C^\mathrm{g}_\mathrm{min}=400$.
}
\end{figure}

The full optimization of all junction areas $\{\mathcal{A}_n\}$, subject to
constraint~(\ref{Ltot=}), can be performed numerically. For any
$N>N_0$, we maximize $\omega_1$ as a function of all the areas,
calculated numerically from the eigenvalue equation
$\det\hat\Upsilon(\omega)=0$. The resulting maximum $\omega_1$
is plotted versus $N$ in Fig.~\ref{fig:areaomega1} for several
values of $C_\mathrm{min}/C^\mathrm{g}_\mathrm{min}$ and
$\mathcal{A}_\mathrm{max}$. 
The analytical result of \ref{app:2central} shows that
the curve starts to bend down at $N=N_0+1$, and the numerics shows
that the same trend is followed for all~$N$. Thus the optimal
$\omega_1$ at $N>N_0$ is always below the best value for the
homogeneous chain, $\Omega_{N_0}$. 
In Fig.~\ref{fig:areaprofile}
we show the optimal spatial profile $\{\mathcal{A}_n\}$, corresponding to
one of the points in Fig.~\ref{fig:areaomega1}. Indeed, the best
$\omega_1$ for a fixed $N$ is obtained by placing the largest
junctions in the middle of the chain. Still, the resulting gain
in $\omega_1$ is smaller than the loss due to the increase of the
chain length from $N_0$ to~$N$.

\section{SQUID loop area modulations}
\label{sec:looparea}

As in the previous section, we start by a straighforward
study of the homogeneous case.
Constraint~(\ref{Ltot=}) fixes $\mathcal{F}=N/N_0$, so
for the chain with $L_n=L_\mathrm{max}N_0/N$,
$C_n=C_\mathrm{min}$, $C_n^\mathrm{g}=C_\mathrm{min}^\mathrm{g}$,
assuming $N\gg{1}$, instead of Eq.~(\ref{OmegaNA=}) we have
\begin{equation}\label{OmegaNF=}
\Omega_N^2\approx
\frac{\pi\lambda}{L_\mathrm{tot}C_\mathrm{min}}\,
\frac{N/(\pi\lambda)}{1+N^2/(\pi\lambda)^2},
\end{equation}
where
$\lambda=\sqrt{C_\mathrm{min}^\mathrm{g}/C_\mathrm{min}}$
is the screening length, which does not depend on $N$ and
$\{\mathcal{F}_n\}$ for problem
(\ref{Ln=Fn=})--(\ref{Fhypercube=}).
Expression (\ref{OmegaNF=}) reaches maximum at $N=\pi\lambda$,
so we have to consider three cases for the position of this
value with respect to the interval~(\ref{Ninterval=}).

(i) In the case $\pi\lambda>\mathcal{F}_\mathrm{max}N_0$,
the frequency is maximized by taking the longest possible
chain, $N=\mathcal{F}_\mathrm{max}N_0$. This case
corresponds to the regime when for all allowed~$N$ the first
mode is on the flat part of the mode dispersion curve
(Fig.~\ref{fig:dispersion}).
This means that we have demanded a value of $L_\mathrm{tot}$
which is too small; a larger inductance can be obtained by
simply increasing the length at almost no cost in $\omega_1$.
So, this case has no practical relevance.

(ii) When
$N_0\leqslant\pi\lambda\leqslant\mathcal{F}_\mathrm{max}N_0$,
the frequency is maximized at $N=\pi\lambda$. This corresponds
to the first mode frequency roughly at the boundary between the
flat part of the mode dispersion curve and its acoustic part.

(iii) In the case $\pi\lambda<N_0$, the frequency is maximized
by taking the shortest possible chain. This regime corresponds
to demanding such a large inductance
$L_\mathrm{tot}$ that the first mode necessarily belongs
to the acoustic part of the dispersion curve.
This is the regime where the competition between
$L_\mathrm{tot}$ and $\omega_1$ is the most severe;
it is in this regime that a gain in $\omega_1$ by introducing
a spatial variation of $\mathcal{F}_n$ would be the most
interesting for practical purposes.

The perturbation theory in small modulations $\Delta\mathcal{F}_n$
with respect to a homogeneous chain with $N$ junctions gives a
result, similar to Eq.~(\ref{perturbative=}):
\begin{equation}
\frac{\Delta\omega_1}{\Omega_N}=
\frac{1}{N+1}\sum\limits_{n=1}^N\frac{\Delta\mathcal{F}_n}{\mathcal{F}}
\sin^2\frac{\pi{n}}{N+1},
\end{equation}
which again tells us that inductance modulations in the center
of the chain contribute the most to the increase in~$\omega_1$.
As in the previous section, we now consider a chain of
length $N=N_0+1$ with inductances of two junctions in the center
smaller by a factor $\mathcal{F}=2$. The explicit calculation
of given in \ref{app:2central} shows that this chain
has $\omega_1>\Omega_{N_0}$, and thus one can indeed improve over
the homogeneous result. However, for long chains, $N_0\gg\pi\lambda$,
the gain is quite small:
\begin{equation}
\omega_1-\Omega_{N_0}\approx\frac{1}{2N_0\sqrt{LC}}
\left(\frac{\pi\lambda}{N_0}\right)^3.
\end{equation}

\begin{figure}
\centering
\includegraphics[width=7cm]{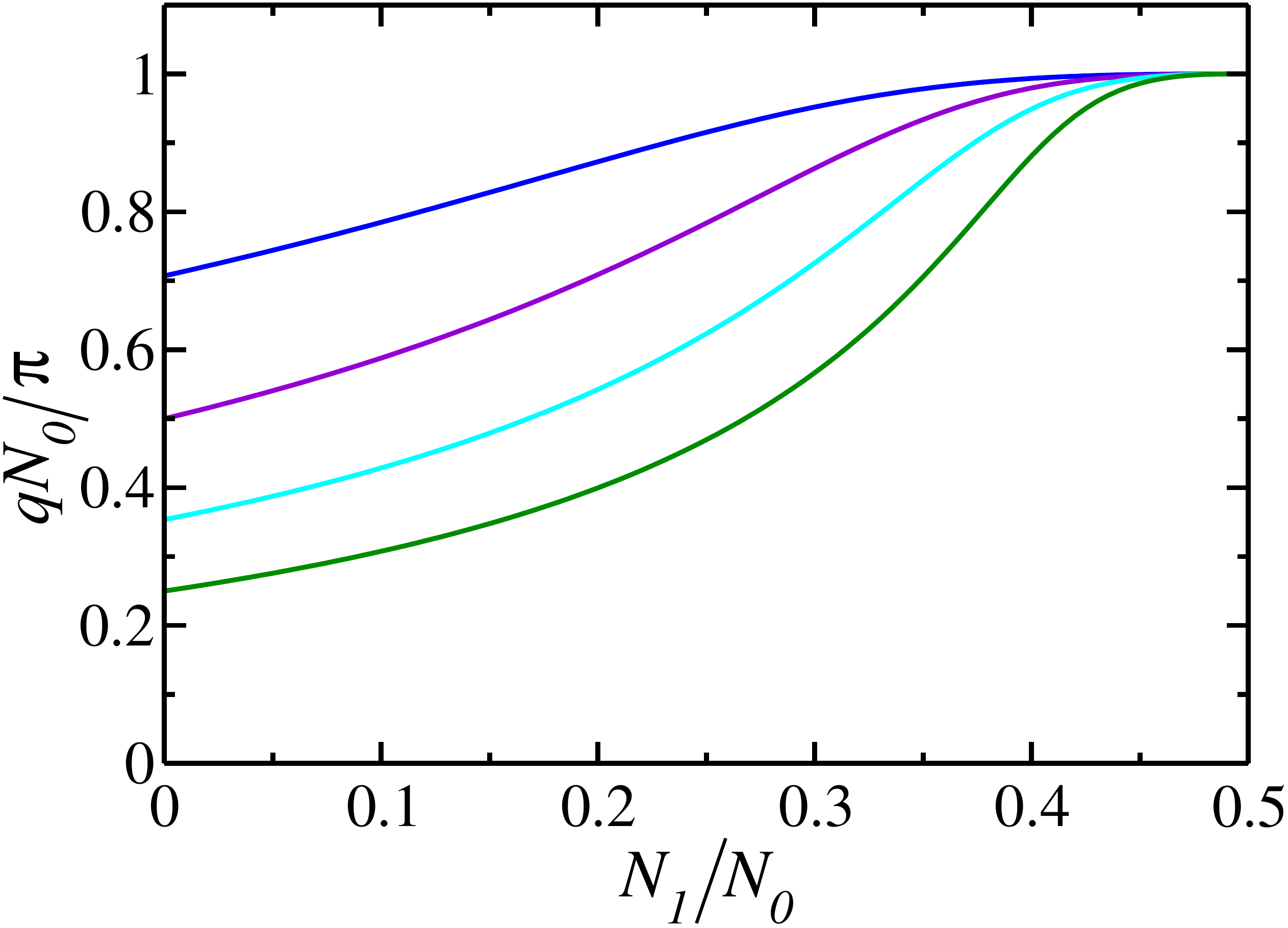}
\caption{\label{fig:continuum}
Solution of Eq.~(\ref{continuum=}) as a function of $N_1$
for different values of $\mathcal{F}=2,4,8,16$ (from the upper to
the lower curve, respectively).
}
\end{figure}

Is it possible to gain more in $\omega_1$ by choosing a chain
length~$N$ significantly exceeding~$N_0$?
As a trial spatial profile, let us consider a long chain with
a central region of length $N-2N_1\gg{1}$ where the inductances
are smaller by a factor $\mathcal{F}$ than in the surrounding
(although this piecewise profile does not coincide with the true
optimal one, found numerically below, it allows for a simple
analytical solution):
\begin{equation}\label{piecewiseL=}
L_n=\left\{\begin{array}{ll}
L_\mathrm{max},&1\leqslant{n}\leqslant{N}_1,\\
L_\mathrm{max}/\mathcal{F},&N_1<n<N-N_1,\\
L_\mathrm{max},&N-N_1\leqslant{n}\leqslant{N}.
\end{array}\right.
\end{equation}
Constraint~(\ref{Ltot=}) then fixes
\begin{equation}
N_1=\frac{\mathcal{F}N_0-N}{2(\mathcal{F}-1)}.
\end{equation}
For $N-2N_1\gg{1}$, we can study the problem in the continuum limit,
replacing the junction number~$n$ by a continuous variable~$x$.
In addition, let us focus on the most interesting case of long chains
$N_0\gg\pi\lambda$, then one can approximate the mode dispersion by
the acoustic one, $\omega(q)\approx{q}/\sqrt{LC^\mathrm{g}}$. Then,
Eqs.~(\ref{Kirchhoff=}) are transformed into the Helmholtz equation
with von Neumann boundary conditions at the ends of the chain,
\begin{equation}
\left(\frac\partial{\partial{x}}\frac{1}{L(x)}\frac{\partial}{\partial{x}}
+\omega^2C^\mathrm{g}\right)V(x)=0,\quad
\left.\frac{\partial{V}}{\partial{x}}\right|_{x=0,N}=0.
\end{equation}
For the piecewise function $L(x)$, given by Eq.~(\ref{piecewiseL=}),
and for a given frequency~$\omega$, the wavenumbers in the outer regions
and in the central region are given by
$q=\omega\sqrt{L_\mathrm{max}C^\mathrm{g}_\mathrm{min}}$ and by
$q/\sqrt{\mathcal{F}}$, respectively. Thus, taking advantage of the
symmetry of $L(x)$ with respect to $x\to{N}-x$, we seek $V(x)$ in the
form (the first mode is odd)
\begin{equation}
V(x)=\left\{\begin{array}{ll}
A\cos{q}x,&0<x<N_1,\\
A'\sin[q(N/2-x)/\sqrt{\mathcal{F}}],&N_1<x<N-N_1,\\
-A\cos(qN-qx),&N-N_1<x<N.
\end{array}\right.
\end{equation}
The requirement of continuity of $V$ and $(1/L)(\partial{V}/\partial{x})$
at $x=N_1,N-N_1$ yields the following equation for~$q$:
\begin{equation}\label{continuum=}
\tan\left[\sqrt{\mathcal{F}}\,q\left(\frac{N_0}2-N_1\right)\right]
=\sqrt{\mathcal{F}}\cot{q}N_1.
\end{equation}
For all $\mathcal{F}>1$, upon increasing $N_1$ from 0 to $N_0/2$
(that is, upon decreasing $N$ from $N_0\mathcal{F}$ to $N_0$),
the solution monotonically rises from
$q=\pi/(N_0\sqrt{\mathcal{F}})$ to $q=\pi/N_0$ (Fig.~\ref{fig:continuum}),
the highest frequency being achieved in the shortest homogeneous chain.
This means that in the limit $N_0\gg\pi\lambda$ the gain in $\omega_1$
is so small that it is not captured by the acoustic approximation.

\begin{figure}
\includegraphics[width=8cm]{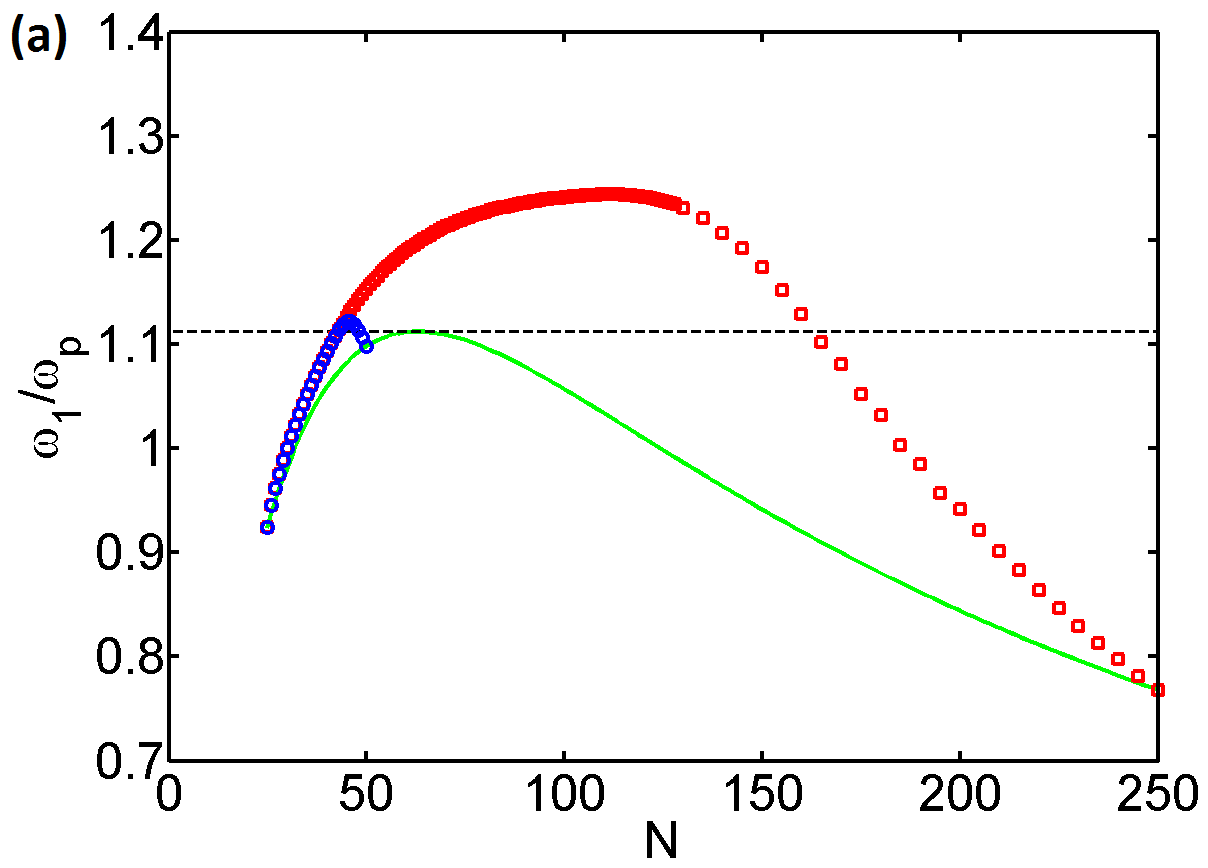}
\includegraphics[width=8cm]{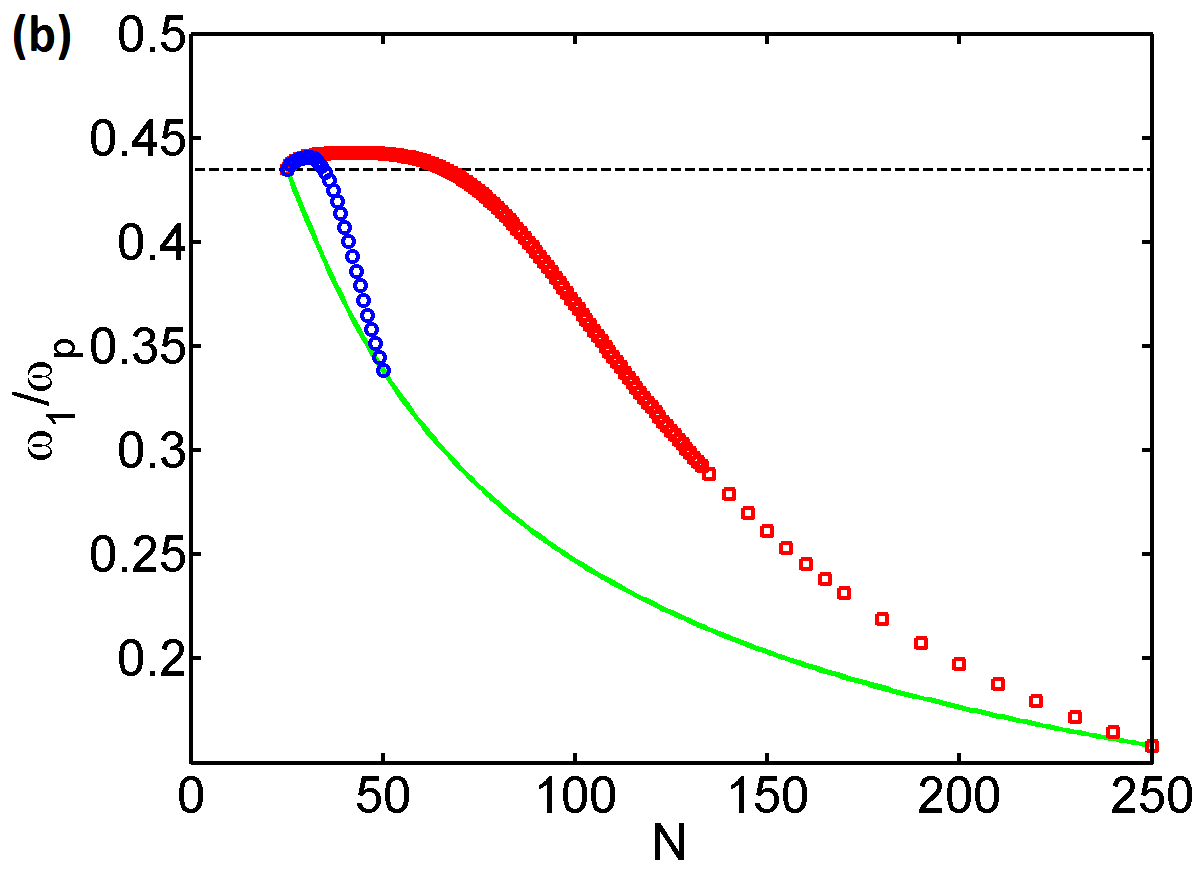}
\caption{\label{fig:fluxomega1}
The first mode frequency~$\omega_1$, in the units of the plasma
frequency
$\omega_\mathrm{p}\equiv{1}/\sqrt{L_\mathrm{max}C_\mathrm{min}}$,
obtained by full numerical optimization of all $\{\mathcal{F}_n\}$,
subject to constraint~(\ref{Ltot=}), shown by symbols for
$\mathcal{F}_\mathrm{max}=2$ and~10 (blue circles and red squares,
respectively). The solid curve shows the first mode
frequency~$\Omega_N$ for the homogeneous chain with
$L_n=L_\mathrm{max}N_0/N$,
$C_n=C_\mathrm{min}$, $C_n^\mathrm{g}=C_\mathrm{min}^\mathrm{g}$.
We take $N_0=25$ for all curves, while
$\lambda=20$ and~4 for panels (a) and~(b), respectively.
The dashed horizontal line shows the best homogeneous
result.
}
\end{figure}

\begin{figure}
\centering
\includegraphics[width=8cm]{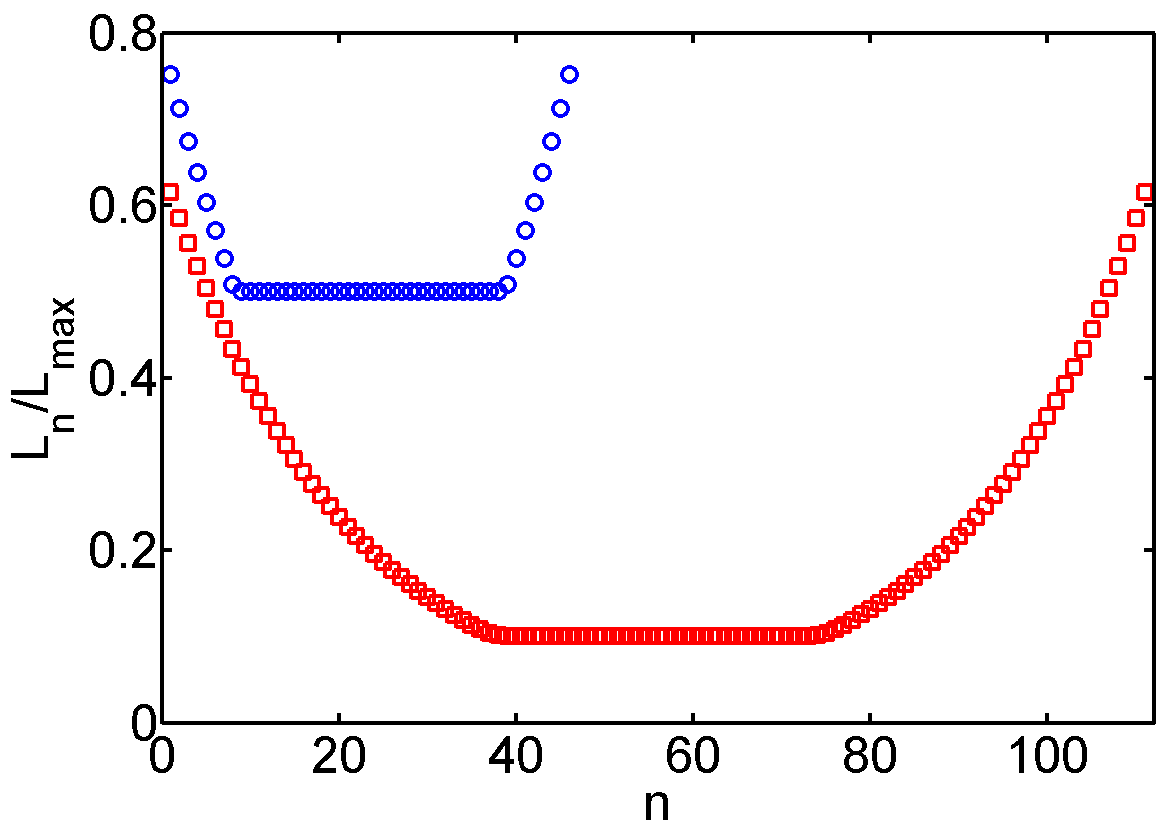}
\caption{\label{fig:fluxprofile}
The optimal spatial profile of the inductance,
$L_n/L_\mathrm{max}$, giving the largest $\omega_1$
for $N=46$, $\mathcal{F}_\mathrm{max}=2$ (blue circles)
and for $N=111$, $\mathcal{F}_\mathrm{max}=10$
(red squares).
Other parameters are $N_0=25$, $\lambda=20$.
}
\end{figure}

\begin{figure}[!htbp]
	\includegraphics[width=8.5cm]{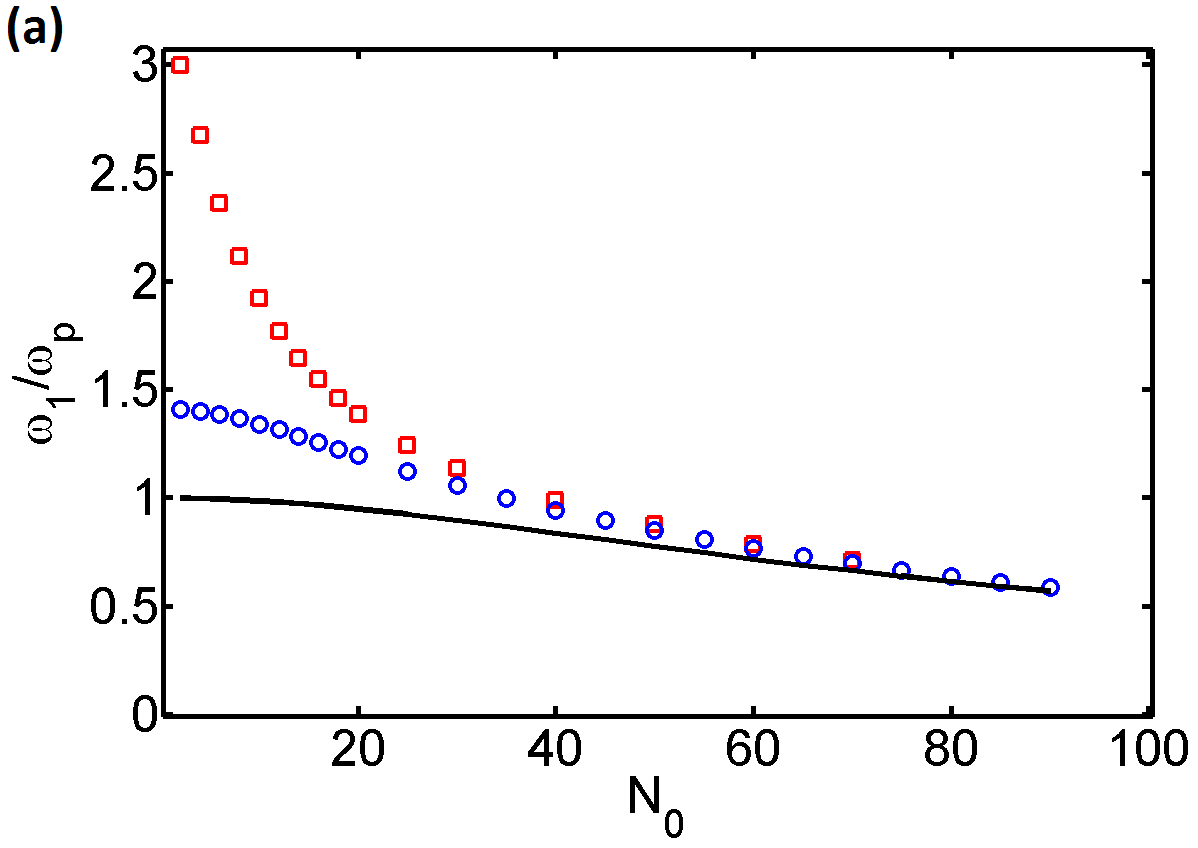} \\
	\includegraphics[width=8.5cm]{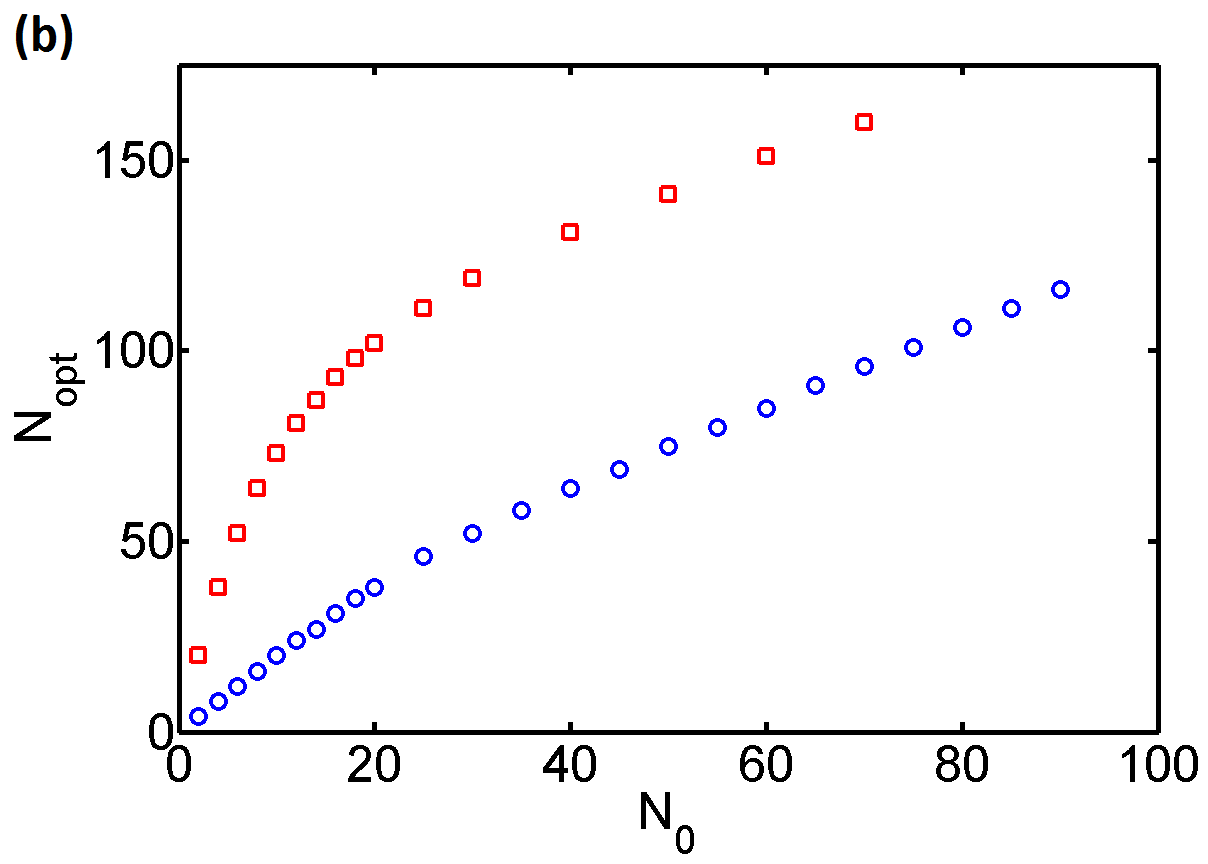}
\caption{\label{fig:longchain}
(a) The $N_0$~dependence of the first mode frequency for
the optimal inhomogeneous chain having the optimal length
(symbols) and for the shortest homogeneous chain with $N=N_0$
(solid curve). The frequencies
are measured in the units of the plasma frequency
$\omega_\mathrm{p}\equiv{1}/\sqrt{L_\mathrm{max}C_\mathrm{min}}$.
(b)~The $N_0$ dependence of the chain length $N_\mathrm{opt}$,
at which the optimal value of $\omega_1$ is obtained.
On both panels the blue circles and the red squares
correspond to $\mathcal{F}_\mathrm{max}=2$ and~10,
respectively, and we took $\lambda=20$.
}
\end{figure}

To check these considerations numerically, we perform the full
optimization of all $\{\mathcal{F}_n\}$, subject to
constraint~(\ref{Ltot=}). As in the previous section, for any
$N>N_0$, we maximize $\omega_1$ as a function of all the areas,
calculated numerically from the eigenvalue equation
$\det\hat\Upsilon(\omega)=0$. The resulting maximum $\omega_1$
is plotted versus $N$ in Fig.~\ref{fig:fluxomega1} for several
values of $\lambda$ and $\mathcal{F}_\mathrm{max}$.
The optimal spatial profile of the inductance is shown in
Fig.~\ref{fig:fluxprofile}; as in the previous section,
it corresponds to putting
the small-inductance junctions in the middle of the chain, and
the large-inductance ones near the ends. From the analytical
arguments above, we do not expect the first mode frequency for
the optimal inhomogeneous chain of optimal length to be much
larger than for the shortest homogeneous chain. This is checked
numerically in Fig.~\ref{fig:longchain}(a), where we plot
the two frequencies as a function of~$N_0$ (we remind that
at fixed~$\mathcal{F}_\mathrm{max}$, $N_0$~parametrizes the
desired total inductance). For long chains, the improvement
due to spatial modulation is indeed negligible. The optimal
length of the modulated chain is close to $N_0$ at large~$N_0$
(up to a constant offset), as shown in Fig.~\ref{fig:longchain}(b).

\section{Conclusions}\label{sec:conclusions}

In this work, we explored the possibility to optimize
the frequency range where a JJ chain can work as
a superinductor, by a careful choice of the spatial
profile of the junction parameters. In the case when
junction areas are varied, the best result is still
obtained for a spatially homogeneous chain, as in
Ref.~\cite{Masluk2012}. Another way to introduce
a spatial variation is to represent the junctions by
SQUIDs whose loop areas are different. Then, by applying
a magnetic field, one can vary the junction inductance
independently from its capacitance. We show that this
strategy can indeed give an improvement with respect
to the homogeneous case, if the most inductive junctions
are placed near the ends of the chain, and the least
inductive ones in the middle. Still, we find that this
improvement becomes less important for longer chains.

{
The qualitative difference between the cases of junction
area and SQUID loop area modulations stems from the fact
that in the first case, the plasma frequency of each
junction, $1/\sqrt{L_nC_n}$, remains fixed. In the second
case, the junction inductance can be decreased independently
from the capacitance, which leads to an increase of the
local plasma frequency, and to a certain degree increases
the overall frequency scale.
}

\begin{acknowledgement}
We are grateful to W.~Guichard and F.~Hekking for
stimulating discussions and critical reading of the
manuscript.
We also acknowledge support from the European Research Council (Grant No. 306731).
\end{acknowledgement}

\setcounter{section}{0}
\renewcommand{\thesection}{Appendix \Alph{section}:}
\setcounter{equation}{0}
\renewcommand{\theequation}{\Alph{section}.\arabic{equation}}

\section{Chain with two central junctions modified}
\label{app:2central}

Let us start by deriving the dispersion relation for
a homogeneous chain of $N$ junctions with parameters
$L_1,\ldots,L_N=L$, $C_1,\ldots,C_N=C$,
$C^\mathrm{g}_1,\ldots,C^\mathrm{g}_{N+1}=C^\mathrm{g}$,
and $\sqrt{C/C^\mathrm{g}}\equiv\lambda$.
Eq.~(\ref{eq:Kirchhoff2}) becomes
\begin{equation}
\frac{1-\omega^2LC}{\omega^2LC^\mathrm{g}}
\left(2{V_n} - {V_{n + 1}} - {V_{n - 1}}\right) - V_n = 0.
\end{equation}
A plane wave, $V_n=A_\pm{e}^{\pm{i}qn}$, with
any $A_\pm$ and $q$ satisfies this equation, provided that
\begin{equation}\label{apd:omega2LC}
\frac{1-\omega^2LC}{\omega^2LC^\mathrm{g}}\,2(1-\cos{q})-1=0,
\end{equation}
which gives the usual dispersion relation~\cite{Masluk2012},
\begin{equation}
\label{apd:dispersion}
\omega(q)  = \frac{1}{{\sqrt {LC} }}
\sqrt{\frac{{1 - \cos q}}{1 - \cos q + 1/(2\lambda^2)}}.
\end{equation}
For a given $\omega$, we seek the solution in the form
$A_+e^{iqn}+A_-e^{-iqn}$, and substitute it into
Eqs.~(\ref{eq:Kirchhoff1}), (\ref{eq:KirchhoffN}) at the ends
of the chain, which play the role of the boundary conditions.
These give, respectively,
\begin{subequations}\begin{align}
& A_+e^{iq}(1-e^{-iq})+A_-e^{-iq}(1-e^{iq})=0,\\
& A_+e^{iq(N+1)}(1-e^{iq})+A_-e^{-iq(N+1)}(1-e^{-iq})=0. 	
\end{align}\end{subequations}
The first of these equations requires the solution to have the
form
\begin{subequations}
\begin{equation}\label{Acos1=}
V_n=A\cos[q(n-1/2)],
\end{equation}
while the second one imposes the form
\begin{equation}\label{BcosN=}
V_n=B\cos[q(n-N-3/2)],
\end{equation}
\end{subequations}
with some $A$ and $B$. Matching these expressions in the
bulk of the chain, we obtain two possibilities, corresponding
to even and odd modes with respect to reflection $n\to{N}+2-n$:
\begin{subequations}\begin{align}
&A=B,\quad q(n-1/2)=q(n-N-3/2)+2\pi{k},\\
&A=-B,\quad q(n-1/2)=q(n-N-3/2)+2\pi{k}+\pi,
\end{align}\end{subequations}
where $k$ is an integer. Thus, the even modes have
$q=2k\pi/(N+1)$, and the odd ones $q=(2k+1)\pi/(N+1)$.
Note that the first mode is odd.

\begin{figure*}[!htbp]
\centering
\includegraphics[width=12cm]{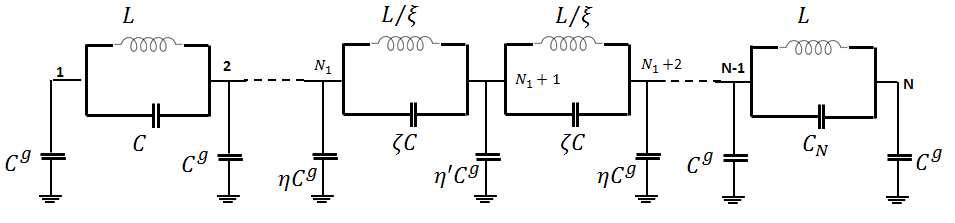}
\caption{\label{fig:apd-area1}
A schematic representation of a JJ chain with two modified
central junctions.}
\end{figure*}

Now, let us consider a chain with two central junctions
modified. We assume $N=N_0+1$ to be even, $N=2N_1$,
then we can again take advantage of the reflection symmetry,
$n\to{N}+2-n$.
In the homogeneous part of the chain we assume
$L_n=L$, $C_n=C$ and $C^\mathrm{g}_n=C^\mathrm{g}$,
while in the central region we set
\begin{subequations}\begin{align}
&L_{N_1-}=L_{N_1+1}=\frac{L}\xi,\\
&C_{N_1}=C_{N_1+1}=\zeta{C},\\
&C^\mathrm{g}_{N_1}=C^\mathrm{g}_{N_1+2}=\eta{C}^\mathrm{g},\quad
C^\mathrm{g}_{N_1+1}=\eta'{C}^\mathrm{g}.
\end{align}\end{subequations}
For the junction areas' variation, considered in
Sec.~\ref{sec:junctionarea}, we have to set
$\xi=\zeta=2$, $\eta=g(3/2)$, $\eta'=g(2)$.
For the loop areas' variation (Sec.~\ref{sec:looparea}),
we have $\xi=2$, $\zeta=\eta=\eta'=1$. The reflection
symmetry is preserved, so the modes can still be classified
as even or odd, and by continuity we know that the first
mode is odd. Thus, similarly to Eqs.~(\ref{Acos1=}),
(\ref{BcosN=}), we can seek $V_n$ in the form
\begin{equation}\label{Vnodd=}
V_n=\left\{\begin{array}{ll}
A\cos[q(n-1/2)],&n\leqslant{N}_1,\\
0,&n=N_1+1,\\
-A\cos[q(n-N-3/2)],&n\geqslant{N}_1+2,
\end{array}\right.
\end{equation}
with yet unknown~$q$ which will be determined by matching
the solutions in the middle of the chain.
Note that as $q$ is related to the frequency
by the dispersion relation~(\ref{apd:dispersion}), which is
a monotonically increasing function, it is sufficient to check
whether the value of~$q$, obtained by matching the solutions,
is larger or smaller than the one corresponding to the shortest
homogeneous chain, $q_0=\pi/(N_0+1)=\pi/(2N_1)$.

$V_n$~in the form~(\ref{Vnodd=}) automatically satisfy the
Kirchhoff laws for the nodes
$n=1,\ldots,N_1-1,N_1+1,N+3,\ldots,2N_1+1$. The Kirchhoff laws
for the remaining $n=N_1,N_1+2$ are identical, so we have one
independent equation which determines~$q$:
\begin{equation}\begin{split}\label{cosqN1=}
&\frac{1-\omega^2LC}{\omega^2LC^\mathrm{g}}
\left[\cos(qN_1-q/2)-\cos(qN_1-3q/2)\right]+{}\\
&{}+\left(\frac{\xi-\zeta\omega^2LC}{\omega^2LC^\mathrm{g}}
-\eta\right)\cos(qN_1-q/2)=0,
\end{split}\end{equation}
where the frequency $\omega$ is related to~$q$ by
Eq.~(\ref{apd:dispersion}).
Note that $\eta'$ dropped out of the equation as $V_{N_1+1}=0$.
Eq.~(\ref{cosqN1=}) can be identically rewritten as
\begin{align*}
&S(q)\cos(qN_1-q/2)-\cos(qN_1-3/2)=0,\\
&S(q)\equiv
1+\xi+\left[\lambda^2(\xi-\zeta)-\eta\right]4\sin^2\frac{q}2,
\end{align*}
or, equivalently, as
\begin{align}
\cot{q}N_1{}
{}={}&{}\frac{2-\xi-[1-\eta+\lambda^2(\xi-\zeta)]4\sin^2(q/2)}%
{\xi+[1-\eta+\lambda^2(\xi-\zeta)]4\sin^2(q/2)}\tan\frac{q}2.
\label{cotqN1=}
\end{align}
The left-hand side of this equation passes through zero precisely
at $q=\pi/(2N_1)=q_0\ll{1}$, with a large negative slope. 
Thus, to find out whether the solution $q=q_*$ is larger or smaller
than $q_0$, we just need to check the sign of the right-hand side
at $q=q_0$.

When only junction areas $\mathcal{A}_{N_1},\mathcal{A}_{N_1+1}$,
are varied, that is, $\xi=\zeta=2$ and $\eta=g(3/2)>1$, the
large factor $\lambda^2$ drops out, so the right-hand side of
Eq.~(\ref{cotqN1=}) is necessarily positive, and thus $q_*<q_0$.
For the variation of SQUID loop areas only,
we have $\xi=2$, $\zeta=\eta=1$, which leads to $q_*>q_0$.
Note, however, that the difference $q_*-q_0$ is quite small:
\begin{equation}
q_*-q_0\approx\frac\pi{2N_0^2}\left(\frac{\pi\lambda}{N_0}\right)^2.
\end{equation}

\end{document}